\documentstyle[preprint,eqsecnum,aps]{revtex}

\begin{document}

\draft

\tighten

\preprint{\vbox{\hfill cond-mat/9809360 \\
          \vbox{\vskip1.0in}
         }}

\title{The approach to thermal equilibrium in quantized chaotic systems}

\author{Mark Srednicki\footnote{E--mail: \tt mark@tpau.physics.ucsb.edu}}

\address{Department of Physics, University of California,
         Santa Barbara, CA 93106
         \\ \vskip0.5in}

\maketitle

\begin{abstract}
\normalsize{
We consider many-body quantum systems that exhibit quantum chaos,
in the sense that the observables of interest act on energy eigenstates
like banded random matrices.  We study the time-dependent
expectation values of these observables, assuming that the system
is in a definite (but arbitrary) pure quantum state.  We induce
a probability distribution for the expectation values by treating
the zero of time as a uniformly distributed random variable.
We show explicitly that if an observable has a nonequilibrium expectation 
value at some particular moment, then it is overwhelmingly likely 
to move towards equilibrium, both forwards and backwards in time.
For deviations from equilibrium that are not much larger than a
typical quantum or thermal fluctuation, we find that the time dependence of
the move towards equilibrium is given by the Kubo correlation function,
in agreement with Onsager's postulate.  These results are independent 
of the details of the system's quantum state.
}
\end{abstract}


\section{Introduction}

Many-body systems typically exhibit certain dynamical properties that
are studied under the subject headings of thermodynamics and
statistical mechanics.  These properties include the following:

1) Given an arbitrary initial state, the system almost always evolves towards
an identifiable condition known as thermal equilibrium, and
then remains in this condition at almost all subsequent times.

2) When the system is in thermal equilibrium, observables
of interest take on values 
that depend only on the nature of the system and its total energy,
but not on any other details of the specific state of the system.

3) When the system is in thermal equilibrium,
the measured value of an observable of interest at any
particular time fluctuates about its equilibrium value,
with fluctuations whose amplitude is suppressed by a factor
of $N^{-1/2}$, where $N$ is the number of degrees of freedom.

4) During the approach to thermal equilibrium,
the values of observables of interest are governed by
equations that are not time reversal invariant.
These equations typically depend on the values of other observables,
possibly at different times.  The information about the system
that is included in these equations is not sufficient to reconstruct 
the complete physical state of the system.

5) Often (but not always), these equations are markovian; 
that is, they depend only on the values of the observables in question,
and their first time derivatives, at any given moment.

There is a vast literature on the derivation of these
properties from an underlying deterministic, time reversal invariant
dynamics, classical or quantum.  In this paper (closely related earlier work
includes \cite{deut,gasp,poil,gv,me,me2,lw,ja,gj,fig,js,tas,pros,tw,silv,cas}),
we explore to what
extent these properties can be deduced as consequences of 
{\em quantum chaos}.  This means that we will assume 
that the energy eigenvalues and (more importantly) the matrix elements 
of typical observables have certain properties.
These properties are believed (and, in some cases, rigorously proven) to hold 
for a canonically quantized system whose classical 
phase space is fully chaotic at the energies of interest,
and they are likely to hold at least approximately
for a broader array of systems.

The outline of this paper is as follows.
In Section 2 we state our basic assumptions, and briefly discuss their
origins in quantum chaos theory.
Some previous work that is directly relevant is summarized in Section 3.  
Sections 4 and 5 present new results concerning the approach to thermal
equilibrium.  Section 6 discusses the main conclusions.

\section{Quantum Chaos}

We assume that the quantum system of interest is bounded and isolated,
with $N$ degrees of freedom, where $N\gg 1$.  Since the system is bounded,
the energy eigenvalues are discrete, and since it is isolated, its time
evolution is governed by the Schrodinger equation.  Let $E_\alpha$
denote the energy eigenvalue corresponding to the energy eigenstate
$|\alpha\rangle$; the state of the system at any time $t$ is then given by
\begin{equation}
|\psi_t\rangle = 
\sum_\alpha c_\alpha\,e^{-iE_\alpha t}\,|\alpha\rangle \,.
\label{psit}
\end{equation}
We emphasize that $|\alpha\rangle$ is an eigenstate of the full,
many-body hamiltonian.
The $c_\alpha$'s specify the state at any one time (say, $t=0$),
and we assume the usual normalization
\begin{equation}
\langle\psi_t|\psi_t\rangle = \sum_\alpha |c_\alpha|^2 = 1 \,.  
\label{norm}
\end{equation}
Note that we have set $\hbar=1$ to simplify the notation.  We will, however,
point out how various quantities scale with $\hbar$ when this information
is useful.

We now make two key assumptions about the system.
  
Our first assumption is not strictly necessary, but it will simplify
some of the subsequent analysis.  We assume that 
if any two sums of equal numbers of energy eigenvalues are equal,
\begin{equation}
E_{\alpha_1} + \ldots + E_{\alpha_n} =
E_{\beta_1} + \ldots + E_{\beta_n} \,,
\label{degen}
\end{equation}
then $\{\beta_1,\ldots,\beta_n\}$ is a permutation of 
$\{\alpha_1,\ldots,\alpha_n\}$; that is, 
the corresponding eigenstates are the same for both sides.
In particular, $E_\alpha = E_\beta$
implies that $|\alpha\rangle = |\beta\rangle$, so that all the energy
eigenvalues are nondegenerate.  This assumption is expected to hold
in general for a quantized chaotic system in which all 
unitary symmetries (such as invariance under any discrete or continuous 
reflection or rotation) have been removed by suitable changes
(such as by putting the system in an irregularly shaped box).
In this case, the energy eigenvalues have the same statistical
properties as the eigenvalues of gaussian random matrices \cite{randmat}.
The system may or may not be invariant under the anti-unitary
transformation of time reversal; we will assume that it is, 
because one of the most interesting aspects of thermodynamics 
is the appearance of an ``arrow of time'' even when the underlying 
dynamics is time reversal invariant.

Our second assumption is the crucial one.  Let $A$ be a hermitian
operator corresponding to an observable of interest that is
a smooth, $\hbar$-independent function of the classical coordinates and
momenta.  We assume that the matrix elements of $A$ in the energy 
eigenstate basis take the form \cite{basic,me2}
\begin{equation}
A_{\alpha\beta} = {\cal A}(E)\delta_{\alpha\beta} 
                  + e^{-S(E)/2} f(E,\omega)R_{\alpha\beta} \,.
\label{aab}
\end{equation}
All the factors in this formula need explanation.  

First, for notational convenience we have defined
\begin{equation}
E \equiv {\textstyle{1\over2}}(E_\alpha+E_\beta) 
\qquad \hbox{and} \qquad
\omega \equiv E_\alpha-E_\beta \,.
\label{eog}
\end{equation}
$S(E)$ is the thermodynamic entropy at energy $E$, given by
\begin{equation}
e^{S(E)} \equiv E \sum_\alpha \delta_\varepsilon(E-E_\alpha) \,,
\label{ent}
\end{equation}
where $\delta_\varepsilon(x)$ is a Dirac delta function that has
been smeared just enough to render $S(E)$ monotonic.
${\cal A}(E)$ and $f(E,\omega)$ are smooth 
functions of their arguments whose physical implications will be the main
focus of this paper.  Finally, $R_{\alpha\beta}$ is a numerical factor that
varies erratically with $\alpha$ and $\beta$.  It is helpful to
think of the real and imaginary parts of $R_{\alpha\beta}$ as random 
variables, each with zero mean and unit variance.  
Without loss of generality, we can take $f(E,\omega)$ 
to be real, positive, and an even function of $\omega$; 
then hermiticity of $A$ implies $R_{\beta\alpha}=R_{\alpha\beta}^*$.
Also, in many cases of physical interest, $R_{\alpha\beta}$ is purely real.

Eq.~(\ref{aab}) is semiclassical in nature; the factor of $e^{-S(E)/2}$ 
scales like $\hbar^{(N-1)/2}$.  Thus the validity of eq.~(\ref{aab}) 
requires $\hbar$ to be ``small,'' which in practice means that the 
energy $E$ must be ``large.''  
The appropriate definitions of 
``small'' and ``large'' are a key problem of quantum chaos theory.  
For systems with few degrees of freedom, 
it is now well established \cite{cc}
that a necessary condition for quantum chaos is 
$\delta \ll \hbar/\tau_{\rm Th}$,
where $\delta \sim e^{-S}E$ is the mean spacing between energy eigenvalues 
near $E$, and $\tau_{\rm Th}$ is the Thouless time \cite{thou}  
(roughly speaking, the time scale 
by which all diffusive classical processes have saturated).
For many-body systems in general, less is known.
Studies of nonlinearly coupled oscillators \cite{lw,tw}
and of fermions with pseudo-random single-particle energies and 
two-body matrix elements \cite{js,silv} both suggest a threshold energy 
for quantum chaos that goes to zero  in the thermodynamic limit like
$N^{-\nu}$, with $0 < \nu \le 1$.  We may therefore hope that 
eq.~(\ref{aab}) will enjoy a wide range of validity.  

An important feature of eq.~(\ref{aab}) is that the general structure
that it describes is preserved under multiplication \cite{me2}.
Thus, for example, the matrix elements of any power of $A$ are given by
\begin{equation}
(A^n)_{\alpha\beta} = {\cal A}_n(E)\delta_{\alpha\beta} 
                  + e^{-S(E)/2} f_n(E,\omega)R^{(n)}_{\alpha\beta} \,,
\label{anab}
\end{equation}
where ${\cal A}_n(E)$, $f_n(E,\omega)$, and $R^{(n)}_{\alpha\beta}$ can
be expressed in terms of 
${\cal A}(E)$, $f(E,\omega)$, and $R_{\alpha\beta}$.
The precise relationship will not be needed, however;
the key point is the generic character of eq.~(\ref{aab}).

Finally, the function ${\cal A}(E)$ can be related to a standard expression
in statistical mechanics:  the equilibrium value of $A$, 
as given by the canonical thermal average
\begin{equation}
\langle A\rangle_T \equiv {\mathop{\rm Tr}e^{-H/T}A \over
                           \mathop{\rm Tr}e^{-H/T}        } \,.
\label{thav}
\end{equation}
Here $T$ is the temperature, and we have set Boltzmann's constant to one.
To see the relation between $\langle A\rangle_T$ and ${\cal A}(E)$,
we use eqs.~(\ref{aab}) and (\ref{ent}) in eq.~(\ref{thav}) to get
\begin{equation}
\langle A\rangle_T = { \int_0^\infty {dE\over E} e^{S(E)-E/T} {\cal A}(E) 
                       \over 
                       \int_0^\infty {dE\over E} e^{S(E)-E/T}             } 
                     + O(e^{-S/2}) \,.
\label{thav2}
\end{equation}
When $N$ is large, the entropy is extensive:
$S(E,N)=Ns(E/N)+O(\log N)$.  In this case
the integrals in eq.~(\ref{thav2})
can be evaluated by steepest descent.  We see that their ratio is
${\cal A}(E) + O(N^{-1})$, where $E$ is now fixed in terms of
$T$ by the steepest-descent condition $\partial S/\partial E = 1/T$;
this also implies $E = \langle H\rangle_T$.
Turning around eq.~(\ref{thav2}) then gives us the expression for
${\cal A}(E)$ that we want \cite{me2},
\begin{equation}
{\cal A}(E) = \langle A\rangle_T + O(N^{-1}) +O(e^{-S/2}) \,.
\label{ae}
\end{equation}
We will always assume that $N$ is large enough to make the indicated
corrections negligible.

Note also that eq.~(\ref{ae}) is consistent with, but not identical to,
Shnirelman's theorem \cite{shnrl}.  This theorem essentially states that
${\cal A}(E)$ is given by the classical, microcanonical average of $A$,
up to corrections which are expected to be $O(\hbar^{1/2})$.
Eq.~(\ref{ae}), on the other hand, already
includes corrections up to $O(\hbar^{(N-2)/2})$, but has
in addition corrections of $O(N^{-1})$.

\section{Thermal Equilibrium}

The expectation value of an observable $A$ in the state specified by
eq.~(\ref{psit}) is given by
\begin{eqnarray}
A_t &\equiv& \langle\psi_t|A|\psi_t\rangle
\nonumber \\
\noalign{\medskip}
    &=& \sum_{\alpha\beta}c^*_\alpha c^{\phantom{*}}_\beta\,
        e^{i(E_\alpha-E_\beta)t} A_{\alpha\beta} \,.
\label{at}
\end{eqnarray}
We will take $A_t$ as the main object of study.  It is not obvious
that this is the right thing to do, since short-time measurements do not 
generally yield quantum expectation values.  However, our main goal is to
compare with the results of conventional nonequilibrium statistical mechanics, 
in which time-dependent expectation values are the basic ingredients
(see, e.g., \cite{kubo,proj,rm}, and Section~V, below).
A detailed discussion of the quantum measurement problem 
would be needed to address this issue properly,
but this is beyond the scope of the present paper.

If we now take the infinite time average of $A_t$, we find
\begin{eqnarray}
\overline{A} &\equiv& \lim_{\tau\to\infty}{1\over \tau}
                      \int_0^\tau dt\; A_t
\nonumber \\
\noalign{\medskip}
             &=& \sum_\alpha |c_\alpha|^2 A_{\alpha\alpha} \,,
\label{ita}
\end{eqnarray}
where the last line requires nondegeneracy of the energy eigenvalues.
It also requires that the averaging time $\tau$ be much larger than
the Heisenberg time $\tau_{\rm H} \equiv 2\pi\hbar/\delta \sim e^S$.  
This time scale is much too large to be physically relevant, and
thus the infinite time average must be regarded as a purely theoretical
device.  Nevertheless, if the system comes to thermal equilibrium, then 
$A_t$ should be near its equilibrium value $\langle A\rangle_T$ 
most of the time, and thus we should have 
$\overline{A} = \langle A\rangle_T.$

To check whether or not this is the case,
we first substitute eq.~(\ref{aab}) into eq.~(\ref{ita}) to get
\begin{equation}
\overline{A} = \sum_\alpha |c_\alpha|^2 {\cal A}(E_\alpha) +O(e^{-S/2}) \,.
\label{ita2}
\end{equation}
We now make a mild assumption about the state $|\psi_t\rangle$.
We assume that the expectation value of the total energy
\begin{equation}
E \equiv \sum_\alpha |c_\alpha|^2 E_\alpha 
\label{eav}
\end{equation}
has a quantum uncertainty 
\begin{equation}
\Delta \equiv \biggl[\,\sum_\alpha |c_\alpha|^2\,(E_\alpha-E)^2\,
              \biggr]^{1/2}
\label{deltae}
\end{equation}
that is small, in the sense that
$\Delta^2 |{\cal A}''(E)/ {\cal A}(E)| \ll 1$.
This is a natural assumption when $N$ is large, since states of physical
interest typically have 
$\Delta \sim N^{-1/2}E$.  
If we now expand ${\cal A}(E_\alpha)$ in eq.~(\ref{ita2})
in powers of $E_\alpha-E$, we get $\overline{A}={\cal A}(E)+O(\Delta^2)$;
combining this with eq.~(\ref{ae}), we find
\begin{equation}
\overline{A} = \langle A\rangle_T +O(\Delta^2) + O(N^{-1}) + O(e^{-S/2})\,.
\label{ita3}
\end{equation}
Thus we have shown that the infinite time average of $A_t$ is indeed
equal to its equilibrium value at the appropriate temperature.
Note that this key property follows entirely from the matrix element
structure of eq.~(\ref{aab}), and does not depend on the details
of the initial quantum state.

We now turn to an examination of the fluctuations of $A_t$ about
its equilibrium value $\overline A$.  The mean squared amplitude
of the these fluctuations is 
\begin{eqnarray}
\overline{ (A_t-\overline{A}\,)^2 }
&=& \lim_{\tau\to\infty}{1\over\tau}
    \int_0^\tau dt\,(A_t-\overline{A}\,)^2 
\nonumber \\
&=& \sum_{\alpha,\beta\ne\alpha} |c_\alpha|^2\,|c_\beta|^2\,
                                              |A_{\alpha\beta}|^2 
\nonumber \\
&=& O(e^{-S}) \,.
\label{deltaita}
\end{eqnarray}
We see that the fluctuations of $A_t$ about $\overline A$ are very small.  
This tells us that, whatever the initial value $A_0$ happens to be, 
$A_t$ must eventually approach its equilibrium value, and then remain
near it most of the time.  

On the other hand, eq.~(\ref{deltaita}) 
is too small to represent the expected thermal fluctuations of $A$.
To find them, we must look at what are usually
called quantum fluctuations.  The mean squared amplitude 
of the quantum fluctuations is
\begin{eqnarray}
\overline{\langle\psi_t| (A-\overline{A}\,)^2 |\psi_t\rangle}
&=& \overline{(A^2)_t} - \overline{A}{}^{\,2}
\nonumber \\
&=& \langle A^2\rangle_T - \langle A\rangle_T^2 
\nonumber \\
&& {} + O(\Delta^2) + O(N^{-1}) + O(e^{-S/2}) \,.
\label{asqita}
\end{eqnarray}
In the last line, we have used the fact that 
the matrix elements of $A^2$ have the same general
structure as the matrix elements of $A$, as shown in eq.~(\ref{anab}),
and that this structure implies that the infinite time average
is the same as the thermal average, as shown in eq.~(\ref{ita3}).
Eq.~(\ref{asqita}) tells us that the
quantum fluctuations in $A$ have the right magnitude to 
be identified as thermal fluctuations \cite{me2}.  

Note, however, that $\langle A^2\rangle_T - \langle A\rangle_T^2$
is itself expected to be $O(N^{-1})$ for 
typical observables of interest (see, e.g., \cite{khin}), 
and so the first two correction terms
on the right-hand side of eq.~(\ref{asqita}) are not necessarily smaller
than the leading term.
This is not a point of concern, however; we used the canonical
ensemble to define thermal averaging, and
the result would in general change by a factor of order one
if we were to use instead the grand canonical or microcanonical ensemble.
Since the exact size of the thermal fluctuations in any particular
observable depends on the choice of ensemble, our claimed identification
of quantum fluctuations as thermal fluctuations can be meaningful only up
to a numerical factor.  This is what is established in eq.~(\ref{asqita}).

Another point of interest is the nature of the classical limit.
Recalling that $e^{-S} \sim \hbar^{N-1}$, we see that
eq.~(\ref{deltaita}) predicts that
$\overline{ (A_t-\overline{A}\,)^2 }$ vanishes in the classical limit.
This is in fact a reasonable result
if the classical system is chaotic.
To see why, first note that the classical
limit of a generic quantum state (to the extent that
such a thing exists at all) is a probability density on phase space.
Then the time dependent expectation value $A_t$ is given 
(in the classical limit) by
$A_t=\int d^{2N}\!\!X\,U_t\rho(X)A(X)$, where $X$ is a point in phase space,
$\rho(X)$ is the probability density 
associated with the quantum state at $t=0$, and
$U_t$ is the Frobenius-Perron
evolution operator for phase-space densities.
When suitably regulated
and renormalized, this formally unitary operator acquires eigenvalues
(known as Ruelle resonances)
inside the unit circle that are associated with irreversible decay
to the ergodic distribution \cite{ruelle}.  
If both $A(X)$ and $\rho(X)$ are
smooth functions, $A_t$ approaches a fixed limit
as $t\to\infty$, and $\overline{ (A_t-\overline{A}\,)^2 }$ vanishes.
The mean squared amplitude of the
classical thermal fluctuations is then given by the infinite time average of
$\int d^{2N}\!\!X\,U_t\rho(X)[A(X)-\overline{A}\,]^2$,
which is the classical limit of the quantum expression
$\langle\psi_t| (A-\overline{A}\,)^2 |\psi_t\rangle$.

\section{Approaching Equilibrium}

We now turn to our main topic, the approach to thermal equilibrium.
Suppose we have an initial state
$|\psi_0\rangle=\sum_\alpha c_\alpha|\alpha\rangle$
such that the initial expectation value
$A_0 = \langle\psi_0|A|\psi_0\rangle$ of an observable $A$
is far from its equilibrium value $\overline{A} = \langle A\rangle_T$.
What, then, is the behavior of 
$A_t = \langle\psi_t|A|\psi_t\rangle$ at later times?

It is obvious that the answer depends on the details of the initial state.
Without knowing them, we can only make a probabilistic analysis.
There are two basic methods for doing so.
One is to introduce a probability distribution for the initial state itself,
and to average relevant quantities over it.
The issue then becomes the justification of the procedure 
(e.g., maximum entropy) for choosing this particular distribution.
The second method, which we will adopt, is to
fix the initial state, and then study the values
of interesting observables as functions of time.  
We treat the observation time as a uniformly distributed random variable,
thus inducing a probability distribution for each observable.
We then attempt to determine to what extent these probability
distributions depend on the initial state.
Ideally, there would be no dependence at all, thus rendering
the choice of the initial state irrelevant.

The moments of the probability distribution for $A_t$ (which is induced 
by assuming a uniform probability distribution for $t$) are 
given by the infinite time averages
\begin{equation}
\overline{ ( A_t \smash{)^n} } = \lim_{\tau\to\infty}{1\over\tau}
                                 \int_0^\tau dt\; (A_t)^n \,.
\label{atn}
\end{equation}
Again we note that we are using the infinite time average simply
as a mathematical tool; in this context, the fact that astronomically long
averaging times are necessary is not relevant.
By using eq.~(\ref{at}), and the nondegeneracy assumption 
discussed after eq.~(\ref{degen}), we can express these
moments as products of traces of powers of the matrix
\begin{equation}
{\widetilde{A}}_{\alpha\beta} \equiv c^*_\alpha A_{\alpha\beta} c_\beta \,.
\label{atilde}
\end{equation}
It will simplify the notation considerably if we first shift 
the operator $A$ by a constant, so that the infinite time
average of $A_t$ is zero.
This entails no loss of generality, and so from here on
we will take 
\begin{equation}
\overline{A}=\mathop{\rm Tr}{\widetilde{A}}=0 \, .
\label{abarzero}
\end{equation}
The first few moments 
$\displaystyle \overline{ ( A_t \smash{)^n} }$ 
can then be expressed as
\begin{eqnarray}
\overline{(A_t)^2} &=&        \mathop{\rm Tr}{\widetilde{A}}^2 \,,
\nonumber \\
\overline{(A_t)^3} &=&     2  \mathop{\rm Tr}{\widetilde{A}}^3 \,,
\nonumber \\
\overline{(A_t)^4} &=&     3 (\mathop{\rm Tr}{\widetilde{A}}^2)^2 
                       +   6  \mathop{\rm Tr}{\widetilde{A}}^4 \,,
\nonumber \\
\overline{(A_t)^5} &=&    20  \mathop{\rm Tr}{\widetilde{A}}^2
                              \mathop{\rm Tr}{\widetilde{A}}^3 
                       +  24  \mathop{\rm Tr}{\widetilde{A}}^5 \,,
\nonumber \\
\overline{(A_t)^6} &=&    15 (\mathop{\rm Tr}{\widetilde{A}}^2)^3
                       +  90  \mathop{\rm Tr}{\widetilde{A}}^2
                              \mathop{\rm Tr}{\widetilde{A}}^4 
                       +  40 (\mathop{\rm Tr}{\widetilde{A}}^3)^2 
                       + 120  \mathop{\rm Tr}{\widetilde{A}}^6    \,.
\label{atns}
\end{eqnarray}
Determining these coefficients is a straightforward but tedious
combinatoric problem; it involves counting the number of different ways 
that the time-dependent phases can cancel against each other, and so
survive the infinite time average.  Generalizing from eq.~(\ref{atns}),
the coefficient of a term of the form
$\prod_i(\mathop{\rm Tr}{\widetilde{A}}^{m_i})^{p_i}$ 
in the expansion of 
$\displaystyle \overline{ ( A_t \smash{)^n} }$ 
is given by 
$n!/(\prod_i p_i! m_i^{p_i})$, where $n=\sum_i m_i p_i$.

We now wish to estimate the magnitude of $\mathop{\rm Tr}{\widetilde{A}}^m$ 
for different values of $m$.  Necessary inputs include the quantum uncertainty
$\Delta$ in the total energy $E$, given by eq.~(\ref{deltae}); 
the energy bandwidth $W$ of the matrix $A_{\alpha\beta}$ near energy $E$, 
given by 
\begin{equation}
W \equiv { \int_{-\infty}^{+\infty} 
            d\omega \, |f(E,\omega)|^2 \over
            |f(E,0)|^2} \,,
\label{w}
\end{equation}
where $f(E,\omega)$ is defined in eq.~(\ref{aab});  
the typical size $a$ of the quantum/thermal fluctuations in $A$, given by
\begin{equation}
a^2 \equiv 
\overline{(A^2)_t} = \sum_\alpha |c_\alpha|^2\,(A^2)_{\alpha\alpha}
                   = \sum_{\alpha\beta} |c_\alpha|^2\,|A_{\alpha\beta}|^2 \,;
\label{a2}
\end{equation}
and the inverse participation ratio
\begin{equation}
{\cal I} \equiv {1\over \sum_\alpha |c_\alpha|^4 } \, ,
\label{ipr}
\end{equation}
which counts the effective number of different
energy levels that are present in the quantum state of the system.
Also, ${\cal I}^{-1}$ can be regarded as the average value of $|c_\alpha|^2$.
Given $\cal I$, it is helpful to define an effective level spacing
between the participating energy eigenstates,
\begin{equation}
\delta_{\rm eff} \equiv {\Delta \over {\cal I}} \, .
\label{deff}
\end{equation}
In general, $\delta_{\rm eff}$ must be greater than or equal to the actual
level spacing $\delta \sim e^{-S}E$.

Before proceeding to evaluate $\mathop{\rm Tr}{\widetilde{A}}^m$,
we must order the various energy scales.
We expect that ``typical'' states of physical interest will have
$\delta_{\rm eff} \sim \delta$ and $\Delta \sim N^{-1/2}E \sim N^{1/2}T$,
since these are properties of a thermal density matrix. 
We also expect that $W$ will be independent of $N$. 
To see why, we turn to the formula for $A_t$, eq.~(\ref{at}).  From it, 
we can see that the time scale for the initial evolution of $A_t$ is either
$\hbar/W$ or $\hbar/\Delta$, whichever is larger.
(Before this time, no relative phases have changed significantly.)
For an observable $A$ of interest, this time scale should be finite
and nonzero in the thermodynamic limit, and
hence independent of $N$.  Since we expect $\Delta \sim N^{1/2}T$, 
it must be $W$ that is independent of $N$.  We therefore conclude that 
\begin{equation}
\delta_{\rm eff} \ll W \ll \Delta  \ll E
\label{wd}
\end{equation}
is the regime of interest.

With these considerations in place, we note that we can regard 
${\widetilde{A}}$ as a banded random matrix of overall size 
$(\Delta/\delta_{\rm eff})\times(\Delta/\delta_{\rm eff})$
and bandwidth $W/\delta_{\rm eff}$.
Within the band of nonzero matrix elements,
the magnitude of a typical entry is
\begin{equation}
{\widetilde{A}_{\rm typ}} \sim {\cal I}^{-1}(W/\delta)^{-1/2}a \,.
\label{attyp}
\end{equation}
This comes from eq.~(\ref{a2}), whose last equality demonstrates
that a typical value of $|A_{\alpha\beta}|^2$ is $(W/\delta)^{-1}a^2$;
the extra factor of ${\cal I}^{-1}$ in eq.~(\ref{attyp}) comes from
the $c_\alpha$'s in eq.~(\ref{atilde}).  We can now estimate 
$\mathop{\rm Tr}{\widetilde{A}}^m$ for even $m \equiv 2k$ as
\begin{eqnarray}
\mathop{\rm Tr}{\widetilde{A}}^{2k}
&\sim& 
\left({\Delta \over \delta_{\rm eff}}\right)
\left({W \over \delta_{\rm eff}}\right)^k
\left({\widetilde{A}_{\rm typ}}\right)^{2k}
\nonumber \\
&\sim& 
\left({\Delta \over \delta_{\rm eff}}\right)
\left({\delta_{\rm eff} \over \Delta}\right)^{2k}
\left({\delta \over \delta_{\rm eff}}\right)^k
a^{2k} \, .
\label{tratn}
\end{eqnarray}
For odd $m\equiv 2k+1$, the expected value of 
$\mathop{\rm Tr}{\widetilde{A}}^{2k+1}$
can be positive or negative; a root-mean-square estimate of its magnitude is
\begin{eqnarray}
\mathop{\rm Tr}{\widetilde{A}}^{2k+1}
&\sim& 
\left({\Delta \over \delta_{\rm eff}}\right)^{1/2}
\left({W \over \delta_{\rm eff}}\right)^k
\left({\widetilde{A}_{\rm typ}}\right)^{2k+1}
\nonumber \\
&\sim& 
\left({\Delta \over W}\right)^{1/2}
\left({\delta_{\rm eff} \over \Delta}\right)^{2k+1}
\left({\delta \over \delta_{\rm eff}}\right)^{(2k+1)/2}
a^{2k+1} \, .
\label{tratn2}
\end{eqnarray}
However, this estimate should be regarded as less trustworthy than
eq.~(\ref{tratn}).

Eqs.~(\ref{tratn}) and (\ref{tratn2})
imply that the probability distribution for $A_t$ has the form
\begin{equation}
P(A_t) \propto \exp\left[
- \left({\Delta \over \delta_{\rm eff}}\right)
F_+ \! \left({A_t\over (\delta/\delta_{\rm eff})^{1/2}a}\right)
- \left({\Delta \over W}\right)^{1/2}
F_- \! \left({A_t\over (\delta/\delta_{\rm eff})^{1/2}a}\right)
\right] \,,
\label{proba}
\end{equation}
where $F_+(x)$ and $F_-(x)$ are even and odd
functions (respectively) whose Taylor expansions
involve purely numerical coefficients.  
Eq.~(\ref{proba}) can be verified by via a Feynman-diagram
expansion for the moments; the quadratic term from $F_+$ provides the
propagator, and the remaining terms (in both $F_+$ and $F_-$) give
the vertex coefficients.

For 
$A_t \mathrel{\raise.3ex\hbox{$<$\kern-.75em\lower1ex\hbox{$\sim$}}}
(\delta/\delta_{\rm eff})^{1/2}a$, we 
can neglect $F_-$ and all but the quadratic term in $F_+$;
we then have
\begin{equation}
P(A_t) \propto \exp\left[ - \, {\xi_2\over2}
                            \left(\Delta\over\delta\right)
                            \left(A_t\over a\right)^{\!2\,}\right] \,,
\label{proba2}
\end{equation}
where $F_+(x)={1\over2}\xi_2x^2+O(x^4)$,
and $\xi_2$ is a number of order one.
Thus, for sufficiently small values
of $A_t$, its probability distribution (as induced by a uniform
distribution for $t$) is gaussian, and independent of the details
of the initial state; only the energy width $\Delta$ of this state
is relevant.  Furthermore, we see again that the fluctuations
of $A_t$ about its infinite time average (which is zero, by construction)
are suppressed by a factor of $(\delta/\Delta)^{1/2} \sim e^{-S/2}$.

For $A_t \mathrel{\raise.3ex\hbox{$>$\kern-.75em\lower1ex\hbox{$\sim$}}}
(\delta/\delta_{\rm eff})^{1/2}a$,
the nonuniversal behavior
of the functions $F_\pm$ becomes relevant.
Initial states that are ``typical'' according to most any
reasonable criterion would all have $\delta_{\rm eff} \sim \delta$.
In this case the nonuniversal corrections are important for 
$A_t \mathrel{\raise.3ex\hbox{$>$\kern-.75em\lower1ex\hbox{$\sim$}}} a$.

So far, our analysis has not addressed our original question:
given the initial value $A_0$, what is the behavior of 
$A_t$ at later times?  To answer this question, we must
compute the conditional probability $P(A_t|A_0)$
to find the value $A_t$ at time $t$, given the value
$A_0$ at time zero.  By the usual rules of probability
theory, this conditional probability can be expressed as
\begin{equation}
P(A_t|A_0) = { P(A_t,A_0) \over P(A_0)}\,,
\label{paa}
\end{equation}
where $P(A_t,A_0)$ is the joint probability 
for the observable $A$ to have the expectation values 
$A_t$ at time $t$ and $A_0$ at time zero.  This joint
probability is induced by assuming a uniform 
distribution for the initial time; hence the 
moments of $P(A_t,A_0)$ are given by
$\displaystyle \overline{ ( A_{t+t'} \smash{)^n} }
               \overline{ ( A_{t'}   \smash{)^m} }$,
where the time averaging is done
with respect to $t'$, with $t$ held fixed.
To compute these moments,
we need an expansion analogous to eq.~(\ref{atns}).
Let us focus on the universal regime.
In this case, the dominant terms in the expansion,
as in eq.~(\ref{atns}), are those with the largest number of traces.  
Keeping only these terms results in a gaussian distribution.
As always for a gaussian distribution, it is completely determined
by its second moments,
\begin{eqnarray}
\overline{A_{t+t'}A_{t'}} &=& 
\sum_{\alpha\beta} |{\widetilde{A}}_{\alpha\beta}|^2 \,                                                 
                        e^{i(E_\alpha-E_\beta)t} \,,
\nonumber \\
\overline{(A_{t+t'})^2} &=& \overline{(A_{t'})^2} = 
\sum_{\alpha\beta}|{\widetilde{A}}_{\alpha\beta}|^2 \,.          
\label{2mom}
\end{eqnarray}
To streamline the notation a little we define a correlation function
\begin{equation}
C(t) \equiv \overline{\vphantom{(}A_{t+t'}A_{t'}} \Big/ 
          \,\overline{(A_{t'})^2}\,.
\label{c}
\end{equation}
Note that, by construction, $C(0)=1$.

These considerations lead to
\begin{equation}
P(A_t,A_0) \propto \exp\left[
- \, {\xi_2 \over 2} \left(\Delta \over \delta\right)
  {(A_t)^2 + (A_0)^2 - 2C(t) A_t A_0
  \over [1-C(t)]^2 a^2 } \right]  \,,
\label{patt}
\end{equation}
which has the correct second moments.
Then, eqs.~(\ref{proba2}), (\ref{paa}), and (\ref{patt}) give us
the conditional probability
\begin{equation}
P(A_t|A_0) \propto \exp\left[
- \, {\xi_2 \over 2} \left(\Delta \over \delta\right)
  {[A_t - C(t) A_0]^2 \over [1-C(t)]^2 a^2 } \right] 
\label{pat0}
\end{equation}
in the universal regime.
Recalling that $\Delta/\delta \sim e^S$,
eq.~(\ref{pat0}) shows us that it is overwhelmingly likely
that we will find $A_t=C(t)A_0$.
In the next section, we will see that the correlation function $C(t)$
does not depend on the quantum state of the system.
Thus, for all practical purposes,
$A_t$ is fully determined just by the initial value $A_0$;
no other information about the state of the system is needed. 

For $A_0 \mathrel{\raise.3ex\hbox{$>$\kern-.75em\lower1ex\hbox{$\sim$}}}
(\delta/\delta_{\rm eff})^{1/2}a$,
the nonuniversal corrections can become important.
We then expect a formula for 
$P(A_t|A_0)$ that is similar to eq.~(\ref{proba}).
Because the large prefactor of
$\Delta/\delta \sim e^S$ should still be present,
we still expect to get an effectively
deterministic equation for $A_t$ as a function
of $t$ and $A_0$, although it will no longer
take the simple form $A_t=C(t)A_0$.

Returning to the universal regime, we need to study the properties of $C(t)$.
This we do in the next section.

\section{Linear Response}

Eqs.~(\ref{2mom}) and (\ref{c}) imply that the correlation function
$C(t)$ is real, even, and has a maximum magnitude of one.
Also, it is quasiperiodic on the scale of the Heisenberg time
$\tau_{\rm H} = 2\pi\hbar/\delta \sim e^S$.  However,
this time scale is much too long to be of physical interest, and so we
can justifiably ignore the quasiperiodicity of $C(t)$.
Then, also using eqs.~(\ref{aab}) and (\ref{atilde}), we get
\begin{eqnarray}
C(t) &\propto& \sum_{\alpha\beta} |c_\alpha|^2 \, |c_\beta|^2 \,
                                  |A_{\alpha\beta}|^2 \,
                                  e^{i(E_\alpha-E_\beta)t} 
\nonumber \\
     &\propto& \int_{-\infty}^{+\infty} 
               d\omega \, |f(E,\omega)|^2 \, e^{i\omega t} \,.
\label{c2}
\end{eqnarray}
The last line shows that $C(t)$ does not depend on the initial state,
and that the bandwidth $W$ of $f(E,\omega)$ sets the
time scale for $C(t)$.

Eq.~(\ref{c2}) can be compared with the results obtained through
more standard methods.
For example \cite{kubo},
we can give $A$ a nonzero expectation value at $t=0$ by 
supposing that, for $t<0$, the system's hamiltonian was perturbed to 
$H+\lambda A$, where $\lambda$ is a constant.
For $t \le 0$, we assume that the quantum state of the
system is described by a thermal density matrix
$\rho_0 \sim e^{-(H+\lambda A)/T}$.
For $t \ge 0$, this state is evolved forward in time
with the original hamiltonian $H$, so that
$\rho_t=e^{-iHt}\rho_0 e^{iHt}$;  
the time-dependent expectation value of $A$ is then
$A_t \equiv \mathop{\rm Tr}\rho_t A$.
To leading order in $\lambda$,
this procedure results in 
$A_t = C_{\rm Kubo}(t)A_0$, where
\begin{equation}
C_{\rm Kubo}(t) \propto \int_0^{1/T}d\mu \, \left\langle
                                        A_H(-i\mu)A_H(t)
                                        \right\rangle_T \,.
\label{kubo}
\end{equation}
Here $A_H(t) \equiv e^{iHt}A e^{-iHt}$ is the operator $A$
in the Heisenberg picture, and the angle brackets
denote canonical thermal averaging as in eq.~(\ref{thav}).
Eq.~(\ref{kubo}) can 
be written in terms of the matrix elements $A_{\alpha\beta}$ as
\begin{eqnarray}
C_{\rm Kubo}(t) &\propto& \sum_{\alpha\beta} {e^{-E_\alpha/T}-e^{-E_\beta/T}
                                          \over
                                          E_\beta-E_\alpha              } \,
                                         |A_{\alpha\beta}|^2              \,
                                         e^{i(E_\alpha-E_\beta)t} 
\nonumber \\
            &\propto& \int_{-\infty}^{+\infty}  d\omega \, 
                                               {\sinh(\omega /2T)
                                                \over
                                                \omega            } \,
                                               |f(E,\omega)|^2 \,
                                                e^{i\omega t} \,.
\label{kubo2}
\end{eqnarray}
This is not the same as eq.~(\ref{c2}).  Suppose, however, that
the bandwidth $W$ of $f(E,\omega)$ is smaller than the temperature $T$,
and that the falloff of $f(E,\omega)$ for $\omega \gg W$ is fast enough
to make the integral in eq.~(\ref{kubo2}) converge.
Then the factor of $\sinh(\omega /2T)/\omega$ will be approximately constant
over the relevant range of $\omega$, and so we will get
\begin{equation}
C_{\rm Kubo}(t) = C(t) + O(W^2/T^2) \,.
\label{kubo3}
\end{equation}
In this case, eq.~(\ref{pat0}) may be viewed as a verification
of Onsager's postulate \cite{on} that a random fluctuation 
(in the value of some quantity) will dissipate in the same way
as an initial value that is produced by an applied force.

Let us now consider the circumstances under which the time evolution
will be markovian.  In the linear regime, and for $t>0$, a markovian equation
has the form $(d/dt)A_t = -\Gamma A_t$, where the parameter
$\Gamma$ must be real and positive.  Given $A_t = C(t)A_0$, this implies
$C(t)=\exp(-\Gamma |t|)$.  From eq.~(\ref{c2}), we see that 
this corresponds to $|f(E,\omega)|^2 \propto 1/(\omega^2+\Gamma^2)$,
and hence $W \sim \Gamma$.
However, $1/(\omega^2+\Gamma^2)$ does not fall off fast enough 
at large $\omega$ for convergence of the integral in eq.~(\ref{kubo2}),
and so we expect some additional suppression of $|f(E,\omega)|^2$ when
$\omega$ is greater than the temperature $T$.
If it happens that $\Gamma$ is much less than $T$,
then the time evolution will still be approximately markovian, 
but with ``memory'' effects on time scales less than $\hbar/T$.
This is consistent with other analyses \cite{green,rm},
which find that the time evolution of expectation values
is always non-markovian on time scales less than $\hbar/T$,
essentially because of the energy-time uncertainty principle.  
The overall conclusion is that markovian time evolution is associated
with an isolated pole in $|f(E,\omega)|^2$ at $\omega = \pm i \Gamma$,
with $\Gamma \ll T$.
However, this structure for $|f(E,\omega)|^2$ is not required by
any general principles, and so non-markovian behavior on all time scales
is an open possibility.

\section{Discussion}

Our major result is eq.~(\ref{pat0}), which gives the probability
to find that an observable $A$ has a quantum expectation value of $A_t$
at time $t$, given that its expectation value is $A_0$ at time zero
(and assuming that $A$ has been shifted, if necessary,
so that the infinite time average of $A_t$ is zero).
This probability is computed for a particular quantum state
$|\psi_t\rangle$, but with the zero of time treated 
as a uniformly distributed random variable.  

To understand the essential features of eq.~(\ref{pat0}),
it is helpful to rewrite it more schematically as
\begin{equation}
P(A_t|A_0) \propto \exp \left\{-\,O(e^S) \left[A_t - C(t) A_0\right]^2/a^2 
                        \right\} \,.
\label{pat1}
\end{equation}
Here $A_t \equiv \langle\psi_t|A|\psi_t\rangle$
is the time-dependent quantum expectation value of $A$;
$a^2$ is the mean squared amplitude of the quantum fluctuations
(which are also to be identified as thermal fluctuations) in $A$;
$S$ is the thermodynamic entropy at energy 
$E=\langle\psi_t|H|\psi_t\rangle$;
and $C(t)$ is the correlation function (normalized to one at $t=0$)
given in eq.~(\ref{c2}), which, under favorable circumstances,
is the same as the Kubo correlation function $C_{\rm Kubo}(t)$
given in eq.~(\ref{kubo2}).  The range of validity of eq.~(\ref{pat1}) is
$A_0 \mathrel{\raise.3ex\hbox{$<$\kern-.75em\lower1ex\hbox{$\sim$}}}
(\delta/\delta_{\rm eff})^{1/2}a$,
where $\delta \sim e^{-S}E$ 
is the mean energy-level spacing near $E$,
and $\delta_{\rm eff} \ge \delta$ is the mean level spacing
for those eigenstates that participate significantly in $|\psi_t\rangle$.
Maximum participation (such as would be predicted by a thermal density
matrix) corresponds to $\delta_{\rm eff} \sim \delta$.

Eq.~(\ref{pat1}) implies an effectively deterministic evolution
equation for $A_t$, given only $A_0$ as input: $A_t = C(t)A_0$.
The probability that this
equation is not satisfied is $O(e^{-S})$.  Since $C(t)$ is an even function
of time, the effective evolution equation is invariant under time reversal.  
However, $C(t) \le C(0)$ for all $t$, and $C(t)$ decays to zero if 
the bandwidth $W$, defined in eq.~(\ref{w}), is finite.
(There will be quasiperiodic resurgences of $C(t)$ on the scale of 
the Heisenberg time  $\tau_{\rm H} = 2\pi\hbar/\delta \sim e^S$, 
but this is much too long to be relevant.)  
Thus a nonzero initial value $A_0$ is overwhelmingly likely to evolve,
both forwards and backwards in time, towards the equilibrium value of zero.
This is of course entirely consistent with the heuristic picture of 
statistical mechanics originally proposed by Boltzmann. 

If the envelope function $|f(E,\omega)|^2$, defined in eq.~(\ref{aab}),
has an isolated pole at $\omega=\pm i\Gamma$,
then the time evolution will be approximately markovian,
with $C(t) \simeq \exp(-\Gamma |t|)$.
This is consistent with 
$C(t)\simeq C_{\rm Kubo}(t)$ only if
$\Gamma \ll T$, where $T$ is the temperature 
corresponding to a thermodynamic energy of $E$.  
There will still be non-markovian
``memory'' effects on time scales less than $\hbar/T$,
which is consistent with the results of other analyses \cite{green,rm}.

To get eq.~(\ref{pat1}) [or, more precisely, (\ref{pat0})],
we had to make a fairly strong assumption about the 
quantum matrix elements of $A$, eq.~(\ref{aab}).
However, this equation is well grounded in quantum chaos theory,
and is likely to be at least approximately valid 
under a fairly wide range of circumstances.

Naturally it would be useful to extend our results to the nonlinear
regime, and naturally this is very much harder to do.
Still we hope to return to this question in future work.  
A combination of our methods 
(which easily demonstrate the ubiquity of thermal equilibrium)
with projection-operator methods (which easily generate equations
for expectation values, assuming appropriately forced thermal density 
matrices as the starting point \cite{proj,rm}) 
might be a fruitful approach.

\begin{acknowledgments}

I am grateful to 
Roland Ketzmerick,
Jim Langer,
and
Doug Scalapino
for helpful comments.
This work was supported in part by NSF Grant PHY--22022.

\end{acknowledgments}

\end{document}